\documentclass{article}
\usepackage{spconf,amsmath}
\usepackage[dvips]{graphicx}
\usepackage{amssymb,color}

\newenvironment{Enumerate}{\begin{enumerate}\setlength{\parsep}{0mm}\setlength{\parskip}{0mm}\setlength{\itemsep}{0.5ex}}{\end{enumerate}}
\newenvironment{Itemize}{\begin{itemize}\setlength{\parsep}{0mm}\setlength{\parskip}{0mm}\setlength{\itemsep}{0.5ex}}{\end{itemize}}

\newcommand{\Bf}[1]{{\bf #1}}
\newcommand{\Cal}[1]{{\cal #1}}

\newcommand{\Funct}[2]{#1\!\left(#2\right)}
\newcommand{\Funcf}[3]{#1\!\left(\frac{#2}{#3}\right)}

\newcommand{\Brace}[1]{\left\{ #1 \right\}}
\newcommand{\Brack}[1]{\left[ #1 \right]}

\newcommand{\AtVal}[2]{\left. {#1} \right|_{#2}}

\newcommand{\Inv}[1]{\frac{1}{#1}}

\newcommand{\Half}{\frac{1}{2}}

\newcommand{\RealS}{\mathbb{R}}

\newcommand{\IntegerS}{\mathbb{Z}}

\newcommand{\argmin}{\mathop{\rm arg\,min}}

\newcommand{\SISum}[1]{\sum_{#1=-\infty}^{\infty}}

\newcommand{\Integ}[4]{\int_{#1}^{#2} #3 \: \mathrm{d}{#4}}

\newcommand{\Deriv}[2]{\frac{\mathrm{d}}{\mathrm{d}{#2}} \, {#1}}

\newcommand{\DPartial}[2]{\frac{\partial^2}{{\partial{#2}}^2} \, {#1}}

\newcommand{\Derivf}[2]{\frac{\mathrm{d}{#1}}{\mathrm{d} {#2}}}

\newcommand{\Erfc}[1]{{\rm erfc}\!\left({#1}\right)}

\newcommand{\Erfcs}[2]{{\rm erfc}\!\left(\frac{#1}{\sqrt{8}{#2}}\right)}

\newcommand{\DefEq}{\stackrel{\text{\tiny def}}{=}}

\begin{document}

\setlength{\abovedisplayskip}{4pt}
\setlength{\belowdisplayskip}{4pt}

\title{OPTIMIZED LEARNED ENTROPY CODING PARAMETERS FOR PRACTICAL NEURAL-BASED IMAGE AND VIDEO COMPRESSION}

\name{Amir Said \qquad Reza Pourreza \qquad Hoang Le\vspace{-3mm}}
\address{Qualcomm AI Research,\sthanks{Qualcomm AI Research is an initiative of Qualcomm Technologies, Inc.} San Diego, CA, USA\vspace{-3mm}}

\maketitle

\begin{abstract}
Neural-based image and video codecs are significantly more power-efficient when weights and activations are quantized
to low-precision integers. While there are general-purpose techniques for reducing quantization effects, large
losses can occur when specific entropy coding properties are not considered. This work analyzes how entropy coding
is affected by parameter quantizations, and provides a method to minimize losses. It is shown that, by using a certain
type of coding parameters to be learned, uniform quantization becomes practically optimal, also simplifying the minimization
of code memory requirements. The mathematical properties of the new representation are presented, and its effectiveness is
demonstrated by coding experiments, showing that good results can be obtained with precision as low as 4~bits per
network output, and practically no loss with 8~bits.
\end{abstract}


\section{Introduction}\label{scIntro}

After years of research showing the advantages of neural-based techniques for video
coding~\cite{Ma:20:ivc,Ding:21:avc}, and recent demonstrations of real-time decoding in mobile
devices~\cite{Qualcomm:21:rtd,Qualcomm:21:rtm}, there is growing interest in their standardization and commercial
deployment~\cite{Ascenso:20:lbi}, and a main practical implementation challenge is the minimization of decoding power
requirements.

Implementing neural network (NN) computations with low-precision integers significantly reduces power needs, but
requires minimizing the effects of quantizing networks weights and activations \cite{Nagel:19:dfq,Nagel:21:awp,Liang:21:paq}.
For neural image and video codecs, efficacy is strongly defined by the entropy coding stage and related networks, and
can be severely degraded.

Analyzing quantization effects in NNs can be difficult because they implement non-linear processes, and learned variables
have unknown properties. Techniques based on re-training under quantization constraints help to reduce the losses, but
have limited efficacy.

In this work it is shown that NN outputs used for entropy coding have well-defined meaning and properties, which can be
used for predicting quantization effects. Furthermore, those properties can be used for modifying the loss function used
while learning, and it is demonstrated that networks can be trained to be minimally sensitive to quantization.

Further analysis shows that this optimized representation is remarkably independent of the quantization, and thus
networks do not need to be retrained even after significant changes to entropy coding design choices. Thus, in those
networks  the uniform quantization can also be extended beyond the hardware requirements, and used to minimize the
memory needed for storing entropy coding tables.

Experimental results show that the proposed solution can guarantee good performance ($\approx$ 2\% redundancy) with
precision as low as 4~bits, and with relative redundancy reduction by a factor of~4 for each additional bit, reaching
less than~0.01\% for 8 bits.

\section{Entropy coding for NN-based codecs}\label{sc:Codecs}

Entropy coding in conventional codecs is based on estimating the probabilities of variables to be coded. For example,
in video standards like HEVC and VVC~\cite{Sullivan:12:hev,Sze:14:ech,Bross:21:vvc}, adaptive binary arithmetic coding
contexts are used for estimating probabilities related to data elements.

\begin{figure}
\centering
\includegraphics[width=75mm]{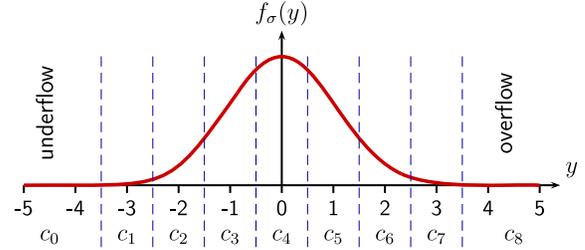}
\caption{\small\label{fg:QuantNormal}Quantized values from normal distribution with std. dev. $\sigma$, commonly used in 
neural-based codecs.\vspace{-4mm}}
\end{figure}

Neural-based codecs use a different principle: a parameterized probability distribution (pdf) is chosen for training,
and a learned scheme (table or NN) is used to determine the parameter value to be used for coding a quantized random
variable. Fig.~\ref{fg:QuantNormal} shows an example of the commonly used normal distribution, and unit-step quantization.

\begin{figure}[t]
\centering
\includegraphics[width=72mm]{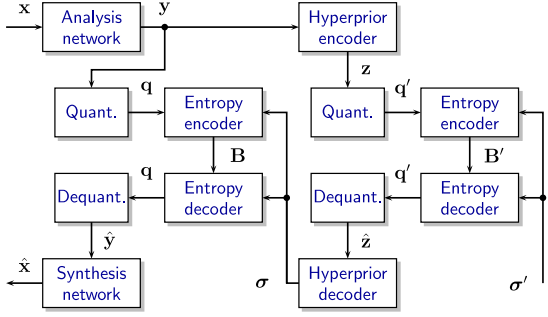}
\caption{\small\label{fg:HyperCodec}Example of a neural-based codec architecture, including scale hyperprior networks for
entropy coding.\vspace{-4mm}}
\end{figure}

The sequence of probabilities of random variable $y \sim \Cal{N}(0,\sigma^2)$, quantized to value $n \in \IntegerS$, is
\begin{equation}
 p_n(\sigma) = \Half \Brack{ \Erfcs{2n - 1}{\sigma} - \Erfcs{2n + 1}{\sigma} }, \label{eq:ProbDefn}
\end{equation}
with entropy
\begin{equation}
 H(\sigma) \DefEq -\sum_{n=-\infty}^{\infty} p_n(\sigma) \Funct{\log_2}{p_n(\sigma)}, \label{eq:Entropy}
\end{equation}
where the {\em complementary error function} is
\begin{equation}
 \Erfc{x} \DefEq \frac{2}{\sqrt{\pi}} \Integ{x}{\infty}{e^{-t^2}}{t}.
\end{equation}

In practical implementations special symbols are used to code values in {\em underflow}
and  {\em overflow} infinite sets, which have very small but nonzero probabilities, as shown in Fig.~\ref{fg:QuantNormal}.
Within those sets the values are sub-optimally coded with, for example, Elias universal codes~\cite{Elias:75:ucs}.

Also shown in Fig.~\ref{fg:QuantNormal} are the elements of {\em code vector $\Bf{c}(\sigma)$} (CV), which contain
the probabilities to be used for arithmetic coding (AC), within a given range $R$
\begin{equation}
 c_i(\sigma) = \begin{cases}
  \sum_{n=-\infty}^{-R} p_n(\sigma), & i = 0, \\
  p_{i-R}(\sigma), & i = 1, 2, \ldots, 2R - 1, \\
  \sum_{n=R}^{\infty} p_n(\sigma), & i = 2R. 
 \end{cases} \label{eq:CVDefn}
\end{equation}

To simplify the notation we ignore implementation details like conversions to integer-valued cumulative sums for
AC~\cite{Said:03:acc,Said:04:iac}. It is assumed that $R$ and AC precision are chosen to obtain compression very close
to entropy.

\begin{figure}[t]
\centering
\includegraphics[width=75mm]{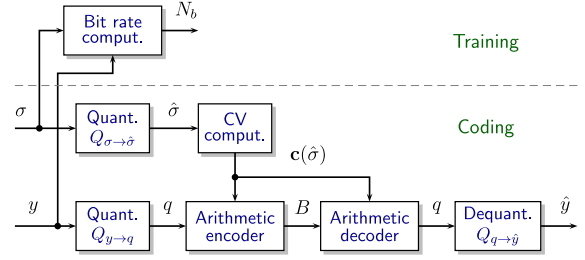}
\caption{\small\label{fg:EntropyCodec}Per-element entropy coding, including quantization of pdf parameter $\sigma$.\vspace{-2mm}}
\end{figure}

This type of entropy coding is used within a NN-based codec as shown in Fig.~\ref{fg:HyperCodec},
corresponding to the scale hyperprior architecture~\cite{Balle:18:vic}. There are other codec
configurations, but our main interest is in the per-element coding stages shown in Fig.~\ref{fg:EntropyCodec},
which is used with all factorized priors.

The latent $y$ to be coded is quantized with $Q_{y \rightarrow q}$, which is taken into account during
floating-point training. The problems addressed here are caused by the quantization stage 
$Q_{\sigma \rightarrow \hat{\sigma}}$, representing the use of low-precision network outputs
during coding, but not during training.

Since code vectors must be bitwise identical at the encoder and decoder~\cite{balle:19:inf}, given a range 
$\Cal{U}=[\sigma_{\min}, \sigma_{\max}]$, we assume they share a monotonic sequence $\{t_k\}_{k=0}^N$
partitioning $\Cal{U}$, and that latent $y$ is coded using $\Bf{c}(\rho^*_k)$ whenever $\hat{\sigma}\in [t_k,t_{k+1})$.

To determine the optimal values of $\rho^*_k$ for code vector computation we measure the {\em average redundancy,}
defined by the Kullback-Leibler divergence
\begin{equation}
 R(\sigma, \rho) \DefEq \sum_{n=-\infty}^{\infty} p_n(\sigma) \Funcf{\log_2}{p_n(\sigma)}{p_n(\rho)}. \label{eq:Redundancy}
\end{equation}

\begin{figure}
\centering
\includegraphics[width=75mm]{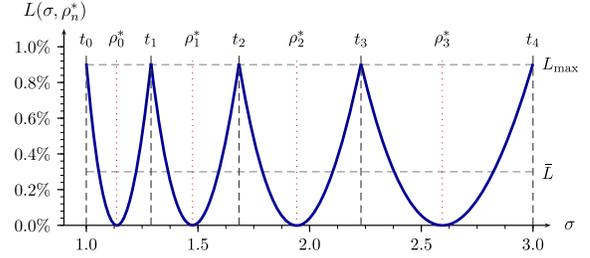}
\caption{\small\label{fg:NUQuant}Example of nonuniform quantization of parameter $\sigma$, designed for constant maximum
relative redundancy.\vspace{-2mm}}
\end{figure}

Since practical image and video codecs are meant to be efficient in a wide range of bit rates, it is interesting to have
entropy coding designed to minimize {\em relative redundancy}
\begin{equation}
 L(\sigma, \rho) \DefEq \frac{R(\sigma, \rho)}{H(\sigma)}. \label{eq:RelRedu}
\end{equation}
and to enable good performance for all use cases use
\begin{equation}
 \rho^*_k = \argmin_{\rho} \Integ{t_k}{t_{k+1}}{L(\sigma, \rho)}{\sigma}. \label{eq:Centroid}
\end{equation}

\section{New pdf parameterization}\label{sc:ModPar}

From these definitions it is possible to optimize non-uniform quantization schemes as shown in the
example of Fig.~\ref{fg:NUQuant}, where similarly to other quantization solutions, all intervals have
the same maximum relative redundancy and averages
\begin{equation}
 \bar{L}_k \DefEq \Inv{t_{k+1} - t_k} \Integ{t_k}{t_{k+1}}{L(\sigma, \rho^*_k)}{\sigma}. \label{eq:AvrgRed}
\end{equation}

With sufficiently fine quantization of $\sigma$, $L(\sigma, \rho^*_k)$ is approximately quadratic, and we can use
\begin{eqnarray}
 \rho^*_k  & \approx & \Brace{ \rho : L(t_k, \rho) = L(t_{k+1}, \rho)}, \label{eq:AppCentroid} \\
 \bar{L}_k & \approx & \frac{L(t_k, \rho^*_k)}{3} =  \frac{L(t_{k+1}, \rho^*_k)}{3}. \label{eq:AppAvrgRed}
\end{eqnarray}

\begin{figure}
\centering
\includegraphics[width=75mm]{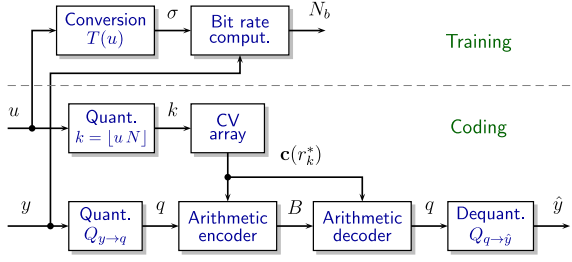}
\caption{\small\label{fg:UniversalCodec}Proposed modification of codec in Fig.~\ref{fg:EntropyCodec}, replacing pdf parameter
$\sigma$ with parameter $u$ optimized for quantization.}
\end{figure}

Neural-based codecs using precise floating-point computations can be implemented with arrays of pre-computed code
vectors, and the approach shown in the example of Fig.~\ref{fg:NUQuant}. However, for
practical codecs, which must handle a wide range of bit rates, it is necessary to support, for example,
$\sigma \in [0.1, 1000]$ with the precision of hyperprior network output $\hat{\sigma}$ constrained to only 8~bits. 
This means that $\{t_k\}_{k=0}^N$ must also use only 8~bits, severely constraining the design and resulting in
significant losses.

\begin{figure}[t]
\centering
\includegraphics[width=75mm]{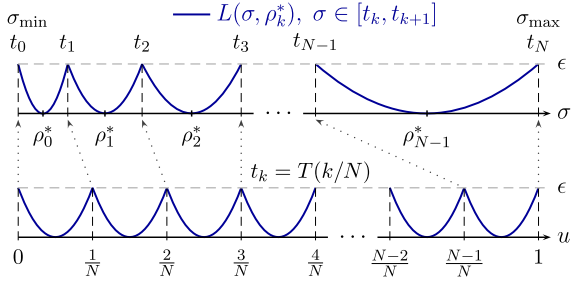}
\caption{\small\label{fg:UMap}Mapping from non-uniform to uniform quantization.}
\end{figure}

To solve this problem we can exploit the fact that $\sigma$ is only used for rate estimation during training, and while
computing code vectors. The proposed solution is shown in Fig.~\ref{fg:UniversalCodec}: hyperprior networks are
trained to compute a new pdf parameter $u \in [0,1]$ that is converted as $\sigma=T(u)$. For analysis we define
$u \in \RealS$, but note that during actual coding the quantized networks only need to generate integer $k$. This approach
is used in~\cite{balle:19:inf}, where the conversion is chosen to be in the form $\sigma=e^{\alpha u + \beta}$.

Fig.~\ref{fg:UMap} shows how optimal $T(u, N)$ can be defined, for a given $N$, to obtain constant maximum
relative redundancy, and the following algorithm can be used for its computation.\vspace{-2mm}
\begin{Enumerate}
 \item Given $N$ and $[\sigma_{\min}, \sigma_{\max}]$, choose initial maximum redundancy $\epsilon$; initialize $t_0=\sigma_{\min}$. 
 \item For $n=1,2,\ldots,N$:
  \begin{Enumerate}
   \item Set $\rho^*_{n-1} = \{\rho > t_{n-1}: L(t_{n-1}, \rho) = \epsilon \}$;
   \item Set $t_n = \{ t > \rho^*_{n-1}: L(t, \rho^*_{n-1}) = \epsilon \}$.
  \end{Enumerate}
  \item If $t_N \approx \sigma_{max}$ then stop.
  \item If $t_N < \sigma_{max}$ then increase $\epsilon$, otherwise decrease $\epsilon$, using a method for unidimensional search.
  \item Go to step~2.
\end{Enumerate}

The surprising conclusion from computing $T(u, N)$ for several values of $N$, and interpolating between the sampled values,
is that all results are practically identical. This means that if we compute 
\begin{equation}
 T(u) = \lim_{N \rightarrow \infty} T(u, N),
\end{equation}
we obtain a single reference function that can be used for all network trainings, and the value of $N$ can be chosen and modified
later, without the need for re-training.

The rationale for determining $T(u)$ can be seen from the lower part of Fig.~\ref{fg:UMap}.
For increasing values of $N$ the redundancy curves in all intervals should approximate quadratic equations on $u$, with same second
derivatives. This can be achieved if we can find $T(u)$ and constant $\alpha$ such that
\begin{equation}
\AtVal{\DPartial{L(T(u), \rho)}{u}}{\rho = T(u)} = \alpha, \quad u \in [0,1]. \label{eq:ConstCurv}
\end{equation}

Defining
\begin{equation}
 \psi(\sigma) \DefEq \AtVal{\DPartial{L(\sigma, \rho)}{\sigma}}{\rho = \sigma} \label{eq:ConstCurve}
\end{equation}
it can be shown that, thanks to properties of $L(\sigma, \rho)$, eq.~(\ref{eq:ConstCurv}) is equivalent to
\begin{equation}
 \Deriv{T(u)}{u} = \sqrt{\frac{\alpha}{\psi(T(u))}}, \label{eq:TODE}
\end{equation}
and thus function $T(u)$ can be determined by solving this ordinary differential equation (ODE) using boundary
conditions $T(0)=\sigma_{\min}, \; T(1)=\sigma_{\max}$.

For any sequence of differentiable parameterized probabilities it can be shown that
\begin{equation}
 \psi(\sigma) = \Inv{\ln(2) H(\sigma)} \SISum{n} \Inv{p_n(\sigma)} \; \Brack{\Derivf{p_n(\sigma)}{\sigma}}^2,
\end{equation}
and for normal pdf we have
\begin{eqnarray}
 \psi(\sigma) & = & \Inv{\ln(2) H(\sigma)} \left\{ \frac{\Brack{\phi_1(\sigma)}^2}{1 - \omega_1(\sigma)} + \right.  \label{eq:LCurvature} \\
 &  & \left. \sum_{n=1}^{\infty} \frac{\Brack{ \phi_{2n-1}(\sigma) - \phi_{2n+1}(\sigma) }^2}{\omega_{2n-1}(\sigma) - \omega_{2n+1}(\sigma)} \right\}, \nonumber
\end{eqnarray}
where
\begin{equation}
 \omega_k(\sigma) \DefEq \Erfcs{k}{\sigma}, \quad \phi_k(\sigma) \DefEq \frac{k \, e^{-k^2/(8\sigma^2)}}{\sqrt{2\pi} \sigma^2}.
\end{equation}

\section{Experimental results}\label{scResults}

\begin{figure}
\centering
\includegraphics[width=75mm]{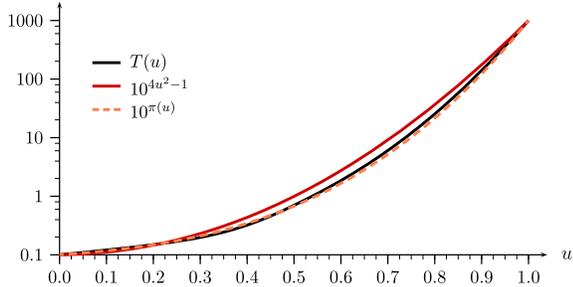}
\caption{\small\label{fg:ConvFunct}The optimal pdf parameter conversion function $T(u)$ and two approximations.}
\end{figure}

\begin{figure}
\centering
\includegraphics[width=75mm]{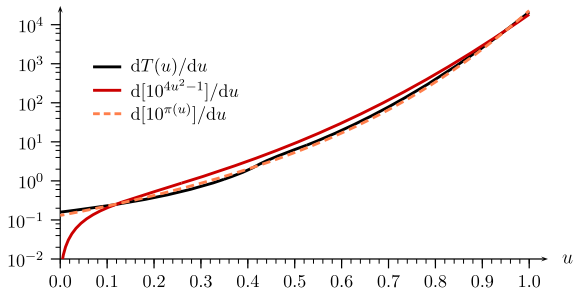}
\caption{\small\label{fg:ConvDeriv}Derivatives of pdf parameter conversion function $T(u)$ and its approximations.}
\end{figure}

The ODE of eq.~(\ref{eq:TODE}) was solved using the 4th-order Runge-Kutta method~\cite[\S17.1]{Press:07:asc}, and boundary
conditions $\sigma \in [0.1, 1000]$. The solution $T(u)$ is shown in Fig.~\ref{fg:ConvFunct}. A quick observation may
lead to the incorrect conclusion that it can be approximated by a function like $10^{4u^2-1}$ (red line).

Since $T(u)$ is defined by a differential equation, the quality of the approximation must be measured by relative
errors of derivatives. Those are shown in Fig.~\ref{fg:ConvDeriv}, where we can also see a much better approximation
$T_{\pi}(u) = 10^{\pi(u)}$, where polynomial 
\begin{equation}
 \pi(u) = 2.49284 \, u^3 + 0.93703 \, u^2 + 0.57013 \, u - 1,
\end{equation}
minimizes the maximum relative derivative error.

\begin{figure}
\centering
\includegraphics[width=75mm]{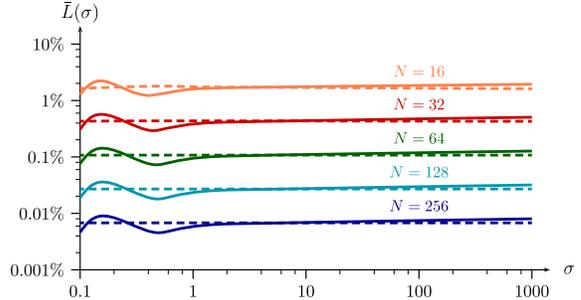}
\caption{\small\label{fg:AvrgRedn}Average relative redundancies from $T(u)$ (dashed) and approximation $10^{\pi(u)}$ (solid), for
different number of quantization intervals $N$.}
\end{figure}

Fig.~\ref{fg:AvrgRedn} shows how the average relative redundancy $\bar{L}(\sigma)$ varies with $\sigma$ when uniform
quantization is applied to pdf parameter $u$, and $N$ code vectors are used for entropy coding latent variables $y$, as
shown in Fig.~\ref{fg:UniversalCodec}. Following the condition defined by eq.~(\ref{eq:ConstCurve}), $\bar{L}(\sigma)$
does not change when $T(u)$ is used for pdf parameter conversion. On the other hand, there are relatively small
variations when approximation $T_{\pi}(u)$ is used instead, and those deviations correspond to errors in 
approximating the derivatives (cf. Fig.~\ref{fg:ConvDeriv}).

We also observe that the curves are nearly perfectly parallel, confirming the predominance of the quadratic term in the
Taylor series of the relative redundancy. This is also verified by the fact that $\bar{L}(\sigma)$ quadruples when the
$N$ is halved.

Non-uniform quantizations defined by transformation $T(u)$ were used for entropy coding values from a floating-point
implementation of the scale hyperprior codec~\cite{Balle:18:vic}, in the Kodak image dataset~\cite{Kodak:set}. Redundancy
results are shown in Table~\ref{tb:CodingResT}, for average bit rates between 0.25 and 2.5~bits/pixel.

For values of $N$ larger than about 128 (7~bit representation) the redundancy from quantization is 0.01\% or below,
which is within the experimental variation. For smaller values of $N$ we can observe that
\begin{Itemize}
\item The redundancies are practically constant for all bit rates, meaning that the proposed coding method achieves the design
 objective of having constant relative redundancies.
\item The average redundancy levels match those of Fig.~\ref{fg:AvrgRedn}, showing that features predicted by theory match
the practical implementation.
\end{Itemize}

\begin{table}
\centering
\begin{footnotesize}
\caption{\small\label{tb:CodingResT}Coding results obtained using scale hyperprior codec applied to 24 images of Kodak dataset.}
\begin{tabular}{|c|c|c|c|c|c|c|} \hline
 $N$ &  CV mem.  & \multicolumn{5}{|c|}{Rel. redundancy (\%) @ bit rate}  \\ \cline{3-7}
     &  (Kbytes) & 0.25 &    0.50 &    1.00 &    1.50 &    2.50 \\ \hline \hline
 256 &   51.6 &    0.00 &    0.00 &    0.00 &    0.00 &    0.01 \\
 192 &   38.7 &    0.00 &    0.00 &    0.00 &    0.01 &    0.01 \\
 128 &   25.8 &    0.04 &    0.04 &    0.04 &    0.03 &    0.03 \\
  96 &   19.4 &    0.03 &    0.04 &    0.04 &    0.04 &    0.05 \\
  64 &   12.9 &    0.13 &    0.12 &    0.12 &    0.12 &    0.11 \\
  48 &    9.7 &    0.24 &    0.23 &    0.22 &    0.22 &    0.20 \\
  32 &    6.5 &    0.38 &    0.38 &    0.40 &    0.41 &    0.42 \\
  24 &    4.9 &    0.85 &    0.81 &    0.81 &    0.82 &    0.79 \\
  16 &    3.4 &    1.79 &    1.76 &    1.76 &    1.78 &    1.75 \\ \hline
\end{tabular}
\end{footnotesize}
\end{table}

The second column of Table~\ref{tb:CodingResT} contains the total amount of memory needed to store the code vectors,
when using 16~bits/element. This shows that the proposed method can be directly used to minimize
memory. For example, even if the network is able to support 8~bit outputs, and redundancy below 0.5\% is acceptable,
the pdf parameter can be further quantized to 5~bits, by simply discarding the 3~least significant bits, to reduce CV memory
from 51.6 to 6.5~Kbytes.

\section{Conclusions}\label{scConclusions}

It is shown that the effects of quantization on the entropy coding can be analyzed by evaluating coding redundancy, and
this can be used for changing the pdf parameter to be learned so that average relative redundancy becomes practically
constant. In consequence, networks do not need to be re-trained when the pdf parameter quantization
is changed, which we show can also be done intentionally to minimize memory for code tables.


\bibliographystyle{IEEEbib}

\begin{thebibliography}{10}

\bibitem{Ma:20:ivc}
S.~Ma, X.~Zhang, C.~Jia, Z.~Zhao, S.~Wang, and S.~Wang,
\newblock ``Image and video compression with neural networks: a review,''
\newblock {\em {IEEE} Trans. Circuits Syst. Video Technol.}, vol. 30, no. 6,
  pp. 1683--1698, June 2020,
\newblock arXiv:1904.03567v2.

\bibitem{Ding:21:avc}
D.~Ding, Z.~Ma, D.~Chen, Q.~Chen, Z.~Liu, and F.~Zhu,
\newblock ``Advances in video compression system using deep neural network: a
  review and case studies,''
\newblock {\em Proc. {IEEE}}, vol. 109, no. 9, pp. 1494--1520, Mar. 2021,
\newblock arXiv:2101.06341v1.

\bibitem{Qualcomm:21:rtd}
Qualcomm AI Research,
\newblock ``World's first software-based neural video decoder running HD format in real-time on a commercial smartphone,''
\newblock {\small URL: https://www.qualcomm.com/news/onq/2021/06/17/worlds-first-software-based-neural-video-decoder-running-hd-format-real-time,} June. 2021.

\bibitem{Qualcomm:21:rtm}
Qualcomm AI Research,
\newblock ``Demonstration of real time neural video decoding on a mobile device,''
\newblock {\small URL: https://www.youtube.com/watch?v=WUnlSHenr08,} Dec. 2021.

\bibitem{Ascenso:20:lbi}
J.~Ascenso, P.~Akyazi, F.~Pereira, and T.~Ebrahimi,
\newblock ``Learning-based image coding: early solutions reviewing and
  subjective quality evaluation,''
\newblock in {\em Proc. SPIE 11353, Optics, Photonics and Digital Tech. for
  Imaging Applicat. {VI}}, Apr. 2020, pp. 164--176.

\bibitem{Nagel:19:dfq}
M.~Nagel, M.~van Baalen, T.~Blankevoort, and M.~Welling,
\newblock ``Data-free quantization through weight equalization and bias
  correction,''
\newblock in {\em Proc. IEEE/CVF Int. Conf. Comput. Vision}, Oct. 2019, pp.
  1325--1334,
\newblock arXiv:1906.04721v3.

\bibitem{Nagel:21:awp}
M.~Nagel, M.~Fournarakis, R.~A. Amjad, Y.~Bondarenko, M.~van Baalen, and
  T.~Blankevoort,
\newblock ``A white paper on neural network quantization,'' arXiv:2106.08295v1,
  June 2021.

\bibitem{Liang:21:paq}
T.~Liang, J.~Glossner, L.~Wang, S.~Shi, and X.~Zhang,
\newblock ``Pruning and quantization for deep neural network acceleration: a
  survey,''
\newblock {\em Neurocomputing}, vol. 461, pp. 370--403, Oct. 2021,
\newblock arXiv:2101.09671.

\bibitem{Sullivan:12:hev}
G.~J. Sullivan, J.-R. Ohm, W.-J. Han, and T.~Wiegand,
\newblock ``Overview of the {High Efficiency Video Coding (HEVC) Standard},''
\newblock {\em {IEEE} Trans. Circuits Syst. Video Technol.}, vol. 22, no. 12,
  pp. 1649--1668, Dec. 2012.

\bibitem{Sze:14:ech}
V.~Sze and D.~Marpe,
\newblock ``Entropy coding in {HEVC},''
\newblock in {\em High Efficiency Video Coding {(HEVC)}: Algorithms and
  Architectures}, V.~Sze, M.~Budagavi, and G.~J. Sullivan, Eds., chapter~8, pp.
  209--274. Springer, 2014.

\bibitem{Bross:21:vvc}
B.~Bross, Y.-K. Wang, Y.~Ye, S.~Liu, J.~Chen, G.~J. Sullivan, and J.-R. Ohm,
\newblock ``Overview of the versatile video coding ({VVC}) standard and its
  applications,''
\newblock {\em {IEEE} Trans. Circuits Syst. Video Technol.}, vol. 31, no. 10,
  pp. 3736--3764, Oct. 2021.

\bibitem{Elias:75:ucs}
P.~Elias,
\newblock ``Universal codeword sets and representations of the integers,''
\newblock {\em {IEEE} Trans. Inf. Theory}, vol. 21, no. 2, pp. 194--203, Mar.
  1975.

\bibitem{Said:03:acc}
A.~Said,
\newblock ``Arithmetic coding,''
\newblock in {\em Lossless Compression Handbook}, K.~Sayood, Ed., chapter~5,
  pp. 101--152. Academic Press, San Diego, {CA}, 2003.

\bibitem{Said:04:iac}
A.~Said,
\newblock ``Introduction to arithmetic coding -- theory and practice,''
\newblock Technical Report HPL-2004-76, Hewlett Packard Laboratories, Palo
  Alto, {CA, USA}, Apr. 2004,
\newblock (http://www.hpl.hp.com/techreports/2004/HPL-2004-76.pdf).

\bibitem{Balle:18:vic}
J.~Ball{\'e}, D.~Minnen, S.~Singh, S.~J. Hwang, and N.~Johnston,
\newblock ``Variational image compression with a scale hyperprior,''
\newblock in {\em Sixth Int. Conf. Learning Representations}, Vancouver,
  Canada, Apr. 2018, arXiv preprint arXiv:1802.01436v2.

\bibitem{balle:19:inf}
J.~Ball{\'e}, N.~Johnston, and D.~Minnen,
\newblock ``Integer networks for data compression with latent-variable
  models,''
\newblock in {\em Int. Conf. Learning Representations}, New Orleans, Louisiana,
  USA, May 2019,
\newblock URL: https://openreview.net/forum?id=S1zz2i0cY7.

\bibitem{Press:07:asc}
W.~H. Press, S.~A. Teukolsky, W.~T. Vetterling, and B.~P. Flannery,
\newblock {\em Numerical Recipes: The Art of Scientific Computing},
\newblock Cambridge University Press, Cambridge, {UK}, third edition, 2007.

\bibitem{Kodak:set}
Kodak Image Dataset,
\newblock URL: {\small http://www.cs.albany.edu/~xypan/research/snr/Kodak.html.}

\end{thebibliography}

\end{document}